\documentclass[10pt, twocolumn, conference]{IEEEtran}

\usepackage{caption}
\usepackage{times,amsfonts,graphicx}
\usepackage[perpage,symbol]{footmisc}
\usepackage[english]{babel}
\usepackage{multirow,multicol}
\usepackage{amsmath} 
\usepackage{pbox, booktabs, rotating}
\usepackage{array, makecell, lipsum}
\usepackage{enumerate}
\usepackage{threeparttable}
\usepackage{multirow,hyphenat,balance}
\usepackage{placeins}
\usepackage[compress]{cite}
\usepackage{float,color}
\usepackage[linesnumbered,ruled]{algorithm2e}
\usepackage{ragged2e}
\usepackage{enumerate}
\usepackage{etoolbox}
\usepackage{gensymb}
\usepackage{rotating, color} 
\usepackage{listings}
\usepackage{booktabs}
\usepackage[dvipsnames]{xcolor}
\usepackage{bm}
\usepackage{boldline}
\usepackage{soul}
\usepackage{adjustbox}
\usepackage{verbatim}
\usepackage{subcaption}

\newcommand{\ineq}[1]{\footnotesize$#1$\normalsize}{}
\usepackage{pifont}
\let\oldding\ding
\renewcommand{\ding}[2][1]{\scalebox{#1}{\oldding{#2}}}

\title{Special Session: Reliability Analysis for ML/AI Hardware}

\author{Shamik Kundu$^1$, Kanad Basu$^1$, Mehdi Sadi$^2$, Twisha Titirsha$^3$, Shihao Song$^3$, Anup Das$^3$, Ujjwal Guin$^2$\\

$^1$Department of Electrical \& Computer Engineering, University of Texas at Dallas (kanad.basu@utdallas.edu)\\
$^2$Department of Electrical \& Computer Engineering, Auburn University (\{mehdi.sadi, ujjwal.guin\}@auburn.edu)\\

$^3$Department of Electrical \& Computer Engineering, Drexel University (anup.das@drexel.edu)
}

\begin{document}
\bstctlcite{IEEEexample:BSTcontrol}

\maketitle

\begin{abstract}
Artificial intelligence (AI) and Machine Learning (ML) are becoming pervasive in today's applications, such as autonomous vehicles, healthcare, aerospace, cybersecurity, and many critical applications. Ensuring the reliability and robustness of the underlying AI/ML hardware becomes our paramount importance. In this paper, we explore and evaluate the reliability of different AI/ML hardware. The first section outlines the reliability issues in a commercial systolic array-based ML accelerator in the presence of faults engendering from device-level non-idealities in the DRAM. Next, we quantified the impact of circuit-level faults in the MSB and LSB logic cones of the Multiply and Accumulate (MAC) block of the AI accelerator on the AI/ML accuracy. Finally, we present two key reliability issues -- circuit aging and endurance in emerging neuromorphic hardware platforms and present our system-level approach to mitigate them.

\end{abstract}

\vspace{5px}

\section{Impact on DRAM Faults on DNN Accelerators}
Deep Neural Networks (DNNs), with their ever increasing computing power are gaining momentum over classical machine learning and computer vision algorithms. As a result, DNNs are being extensively deployed across real-time data driven applications on resource constrained Internet-of-Things (IoT) edge devices. Conventional CPU architectures tend to lose out on implementing the computational complexity of state-of-the-art DNNs, which propelled the development of cost and energy efficient application-specific neural network accelerators. Both industry and academia responded to this emerging community of DNN accelerators by developing several purpose-built inference hardware \cite{jouppi2017datacenter,chen2016eyeriss,park20154}. Google's Tensor Processing Unit (TPU) is one such accelerator that achieves 15–30$\times$ higher performance and 30–80$\times$ higher performance-per-watt over traditional CPUs and GPUs \cite{jouppi2017datacenter}. These custom built DNN accelerators find their application in the domains of computer vision, multimedia processing, graph analytics, and search. With this widespread proliferation of DNNs, researchers have focused on performance optimization and energy-efficiency of such hardware architectures. Even though DNNs are presumed to be resilient against errors by virtue of their inherent fault tolerant capabilities, the threshold of such resiliency can be easily inflicted with bit-level hardware faults, leading to graceless degradation in classification accuracy of the DNN \cite{zhang2018analyzing}.

\begin{figure}[t!]
    \centering
    \includegraphics[width=\linewidth]{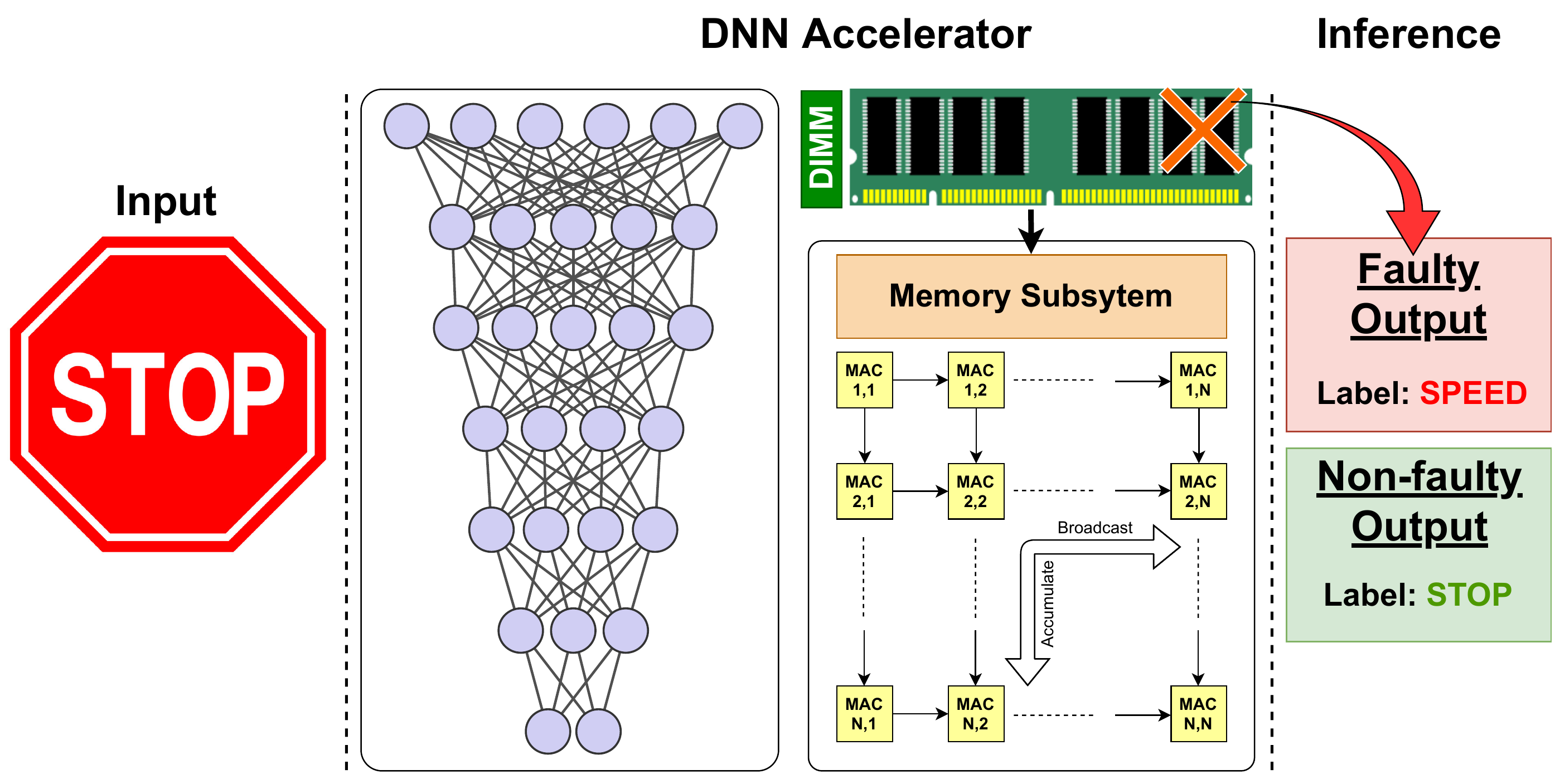}
    \caption{Fault manifestation in DRAM leading to misclassification in a mission-critical DNN accelerator.}
    \label{fig:overview_DRAM}
    \vspace{-0.25in}
\end{figure}

To execute an inference network at the edge, considerable amount of on-chip memory is required to store millions of trained network parameters, input activations and intermediate feature maps. For this purpose, energy restricted DNN accelerators utilize Dynamic Random Access Memory (DRAM) as their primary memory subsystem, owing to its low access latency and high density architecture. However, due to various non-idealities associated with the access transistor, the charge stored in the DRAM cells leak away over time, engendering bit-level faults throughout the memory. In order to alleviate this, a DRAM is periodically replenished using an implicit background refresh operation, which contributes to a significant amount of DRAM energy overhead as well as performance bottleneck in the DNN accelerator. Existing research to alleviate this high refresh rate of DRAMs have focused on utilizing sub-optimal refresh rates; at the expense of bit-flip faults that are highly dependent on temperature and variable retention times of each cell in the structure. Such faults manifested in the structure adversely impact the classification of the network, as demonstrated in Figure~\ref{fig:overview_DRAM}. Since DNN accelerators are often deployed in high-assurance environments, \textit{e.g.}, self driving vehicles for enhancing the autonomous driving dynamics, smart sensors in aerial surveillance vehicles like drones, analyzing the impact of these DRAM faults is highly imperative to avoid catastrophic circumstances.

\subsection{DRAM Basics}

DRAM is the primary choice for main memory in most computing platforms ranging from large scale data centers to miniature edge devices, due to its high density, longevity, and low cost. The basic element of the 2D-DRAM array is a DRAM bitcell, which consists of a single capacitor and an access transistor. Depending on the charge in the DRAM cell capacitor, a binary data value is identified as 1 when the capacitor is fully charged, or 0 when it is in a discharged state. However, DRAM capacitor loses charge over time due to various factors predominantly related to the non-ideality of the access transistor like sub-threshold leakage and gate-induced drain leakage. As a result of this, faults in the form of bit flips start to manifest in the data bits after a certain time interval (known as retention time). Therefore, the charge needs to be refreshed periodically to preserve the integrity of the data. The leakiest cell in the entire DRAM array determines the worst-case retention time, which is usually 64ms in case of commercial DRAM modules. Even though high refresh rates are imperative to maintain correctness, they also adversely impact on the overall energy efficiency and performance. Energy efficiency declines due to the periodic activation of individual rows as well as the increase in energy consumption due to a longer execution time.

Existing research to alleviate the high refresh rate of DRAMs have focused on utilizing sub-optimal refresh rates, approximating the device performance. One of the earliest work in this domain, Flikker, partitions the application data into critical and non-critical parts, injecting errors in the non-critical portion at higher refresh intervals \cite{liu2011flikker}. Sparkk, a DRAM approximation framework refreshes the most significant bits at a higher refresh rate than the least significant ones \cite{lucas2014sparkk}. Other DRAM approximation schemes have also been proposed, that reduces DRAM energy consumption at lower refresh rates \cite{sampson2011enerj, rahmati2014refreshing, raha2017synergistic}. A quality configurable approximate DRAM has been proposed, that utilizes the concept of critical and non-critical data partitioning to allocate data in multiple quality bins, considering the property of variable retention times exhibited by DRAM cells \cite{raha2016quality}. However, the impact of the errors on high assurance DNN accelerators has not been well explored at high DRAM refresh intervals, which drives us to analyze the reliability of such high assurance architectures.

\subsection{DNN Accelerator and Its Reliability}

In recent years, DNNs have acquired a meteoric rise in various spheres of life due to their use of sophisticated mathematical modeling to process data with high complexity parameters. To meet the extreme compute requirements of these compute-heavy DNN algorithms, a number of DNN accelerators have emerged over the past decade. Most DNN accelerators in practice utilizes DRAM as the main memory subsystem. For example, well known DNN accelerators such as Google TPU \cite{jouppi2017datacenter}, Eyeriss \cite{chen2016eyeriss}, NVIDIA Jetson Dev Board \cite{cass2020nvidia}, Google Coral Edge TPU \cite{cass2019taking}, and Intel MyriadX VPU \cite{IntelMo7:online} consists of 8 GB, 1 GB, 4 GB, 1 GB, and 2 GB of DRAM, respectively. Recently, a new class of sparse DNN accelerators has emerged, such as Eyeriss v2 \cite{chen2019eyeriss}, NVIDIA Ampere \cite{NVIDIAAm57:online}, Intel Keem Bay VPU \cite{IntelKeemBayVPU}, \emph{etc.}, that accelerates the performance of sparse matrix convolution in the inference. These accelerators leverage the sparsity in a tensor graph and therefore, skip certain computations during the inference, based on the bitmap encoding of the tensors. In these sparse accelerators, DRAM reliability is extremely critical, which can lead to subverting the accuracy of the DNN accelerator. 

With the extensive deployment of DNN accelerators in a wide gamut of applications, researchers have analyzed the impact of faults in various inference hardware. To explore the correlation between bit error rate and model accuracy, permanent faults are injected in the memory elements of a customized DNN accelerator \cite{reagen2018ares}. The impact of single bit soft errors in the memory on the network performance is explored in \cite{schorn2019efficient}, which further proposed a bit-flip resilience optimization method for DNN accelerators. Memory faults are induced in the Autonomous Driving System (ADS) of a vehicle and the corresponding resilience of different modules are examined \cite{jha2019ml}. The susceptibility of the architecture under single event upsets on the datapath of an accelerator is analyzed on multiple Convolutional Neural Network (CNN) models \cite{li2017understanding}. The safety critical bits in a machine learning system were identified by inducing soft errors in the network datapath and evaluating them on eight DNN models across six datasets \cite{chen2019binfi}. A formal analysis on the impact of faults in the datapath of an accelerator has been illustrated using Discrete-Time Markov Chain (DTMC) formalism \cite{kundu2020high}. The intense performance penalty in a systolic array-based DNN accelerator has been demonstrated by inducing manufacturing defects in the datapath of the accelerator \cite{zhang2018analyzing, zhang2019fault, kundu2021toward}. However, fault characterization of DNN accelerators due to device-level non-idealities in  the DRAM cells has not been well explored.


\subsection{Fault Characterization to Estimate Network Performance}

In order to analyze the impact of DRAM faults on the performance of the DNN accelerator, we implement Multilayer Perceptron (MLP) on two different datasets \textemdash MNIST and Fashion-MNIST. The detailed network configuration of the MLP is provided in Table~\ref{table:mlp}. We consider Google's Tensor Processing Unit as the baseline DNN accelerator, having 8GB of dual-channel DDR3 DRAM as the main memory subsystem. The trained weights from each network are extracted and quantized to 8 bits to be stored in the DRAM, similar to \cite{zhang2018analyzing}. Subsequently, the trained weights from the DRAM are mapped on to the inference hardware. MNIST furnishes a baseline classification accuracy of 97.28\%, whereas Fashion-MNIST manifest an accuracy of 88.17\% on MLP. Errors in the form of bit-flips are induced throughout the DRAM structure, following which the application-level fault characterization of the accelerator is analyzed in this section.

\begin{table}[h!]
\caption{Overview of the MLP Architecture.}
\centering
\begin{tabular}{|c|c|l|}
\hline
\textbf{Dataset}      & \multicolumn{2}{c|}{\textbf{Model Configuration}} \\ \hline \hline
MLP on MNIST  & \multicolumn{2}{c|}{$784-256-256-256-10$}           \\ \hline
MLP on Fashion-MNIST              & \multicolumn{2}{c|}{$784-256-256-256-10$}          \\ \hline
\end{tabular}
\label{table:mlp}
\vspace{-0.1in}
\end{table}

\subsubsection{Impact of Faults for Varying Bit Positions}
\label{subsec:bitpos}

In this experiment, the vulnerability of the network is analyzed for varying bit positions of the induced fault in the 8-bit weight stored in the DRAM. Bit-flips are introduced at random positions and the average classification accuracy of the model is observed for 10 runs. Bit flip faults are induced in the three most significant bit positions, starting from the sign bit.

\begin{figure}
\centering
\begin{subfigure}{0.25\textwidth}
  \includegraphics[width=\linewidth]{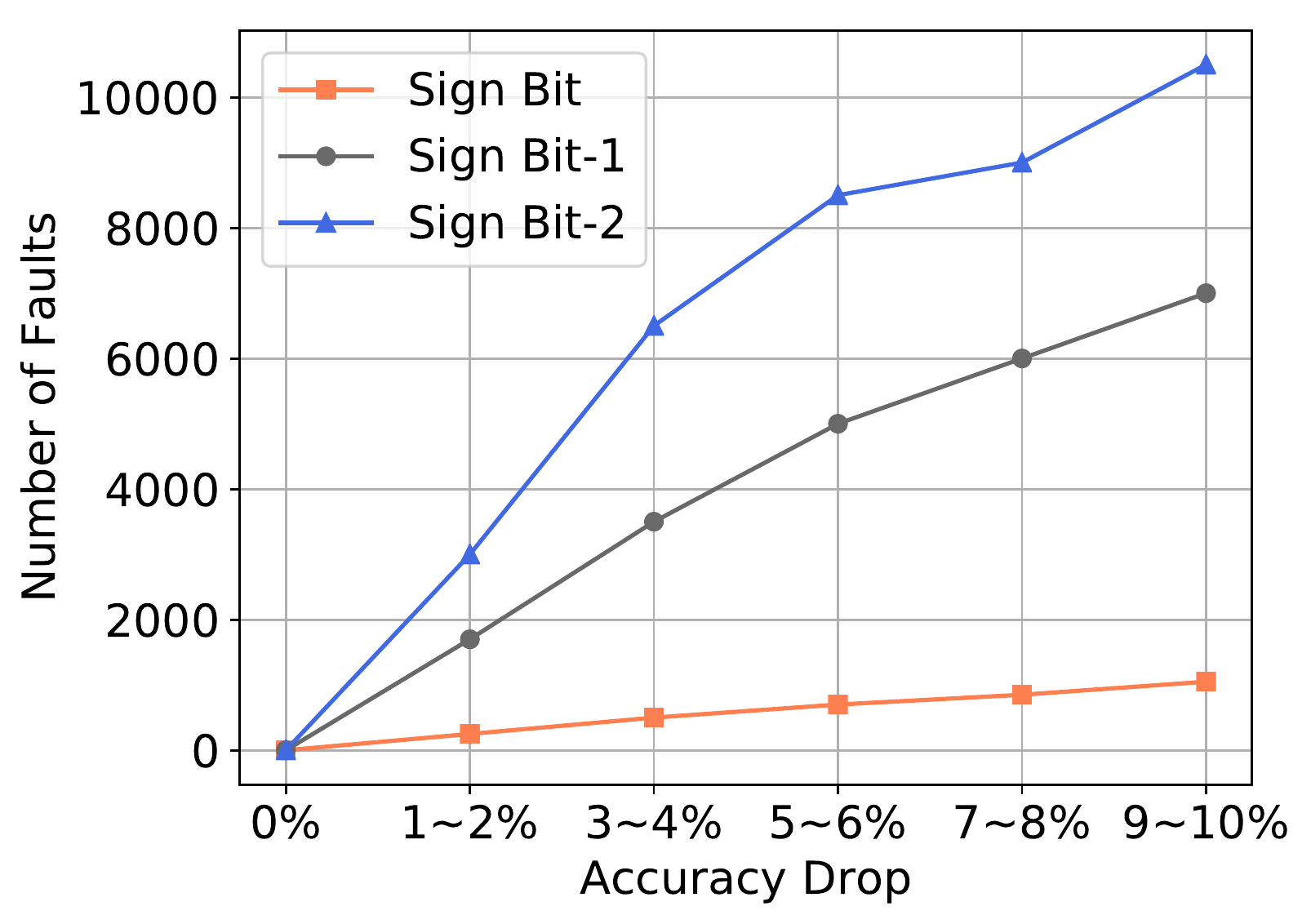}
  \caption{}
  \label{fig:mnist}
\end{subfigure}%
~
\begin{subfigure}{0.25\textwidth}
  \includegraphics[width=\linewidth]{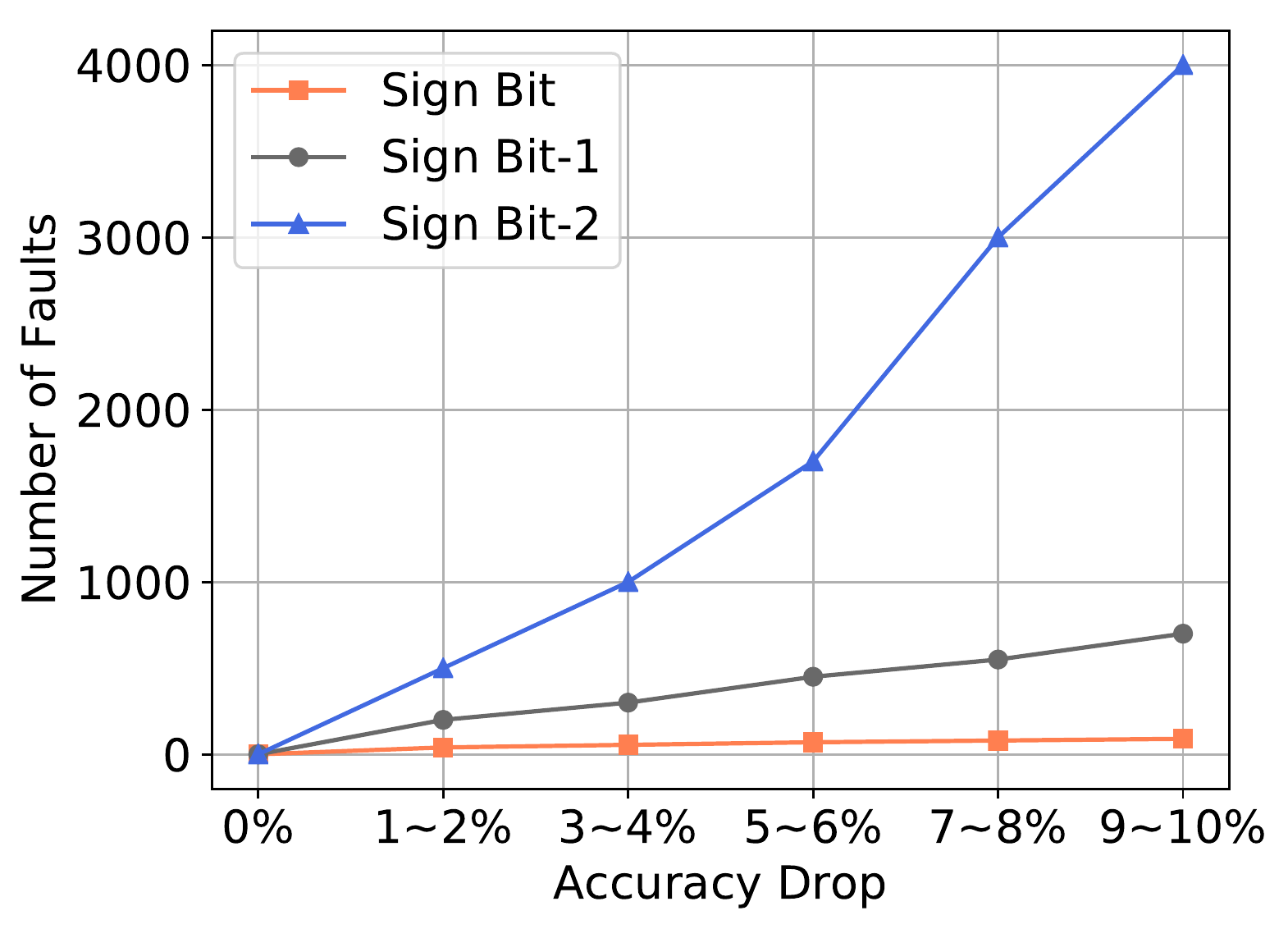}
  \caption{}
  \label{fig:fmnist}
\end{subfigure}%

\caption{Impact of Faulty Bit Position on Classification Accuracy of MLP on (a) MNIST and (b) Fashion-MNIST.}
 \label{fig:DRAMall}
\vspace{-0.2in}
\end{figure}

The corresponding accuracy drop for MLPs on MNIST and Fashion-MNIST are represented in Figures~\ref{fig:mnist} and ~\ref{fig:fmnist} respectively. As seen from the figures, with the increase in number of faults, the classification accuracy of the network reduces for every bit position. As the significance of the bit position diminishes, the sensitivity of the induced fault reduces; thereby increasing the number of faults to accomplish identical reduction in classification accuracy. Since bit-flips induced in the sign bit reverses the signed integer representations, it causes maximum impact on the classification accuracy of the network. Hence, MLPs on MNIST and Fashion-MNIST manifest an 1-2\% reduction in accuracy with only 250 and 40 faults injected in each layer, respectively, at the sign bit of the weights. Thus, minimal faults in the sign bit have the most intense impact on the reliability of the DNN accelerator. 

\subsubsection{Impact of Faults on the Most Vulnerable Weights}

\begin{figure}[t!]
    \centering
    \includegraphics[width=\linewidth]{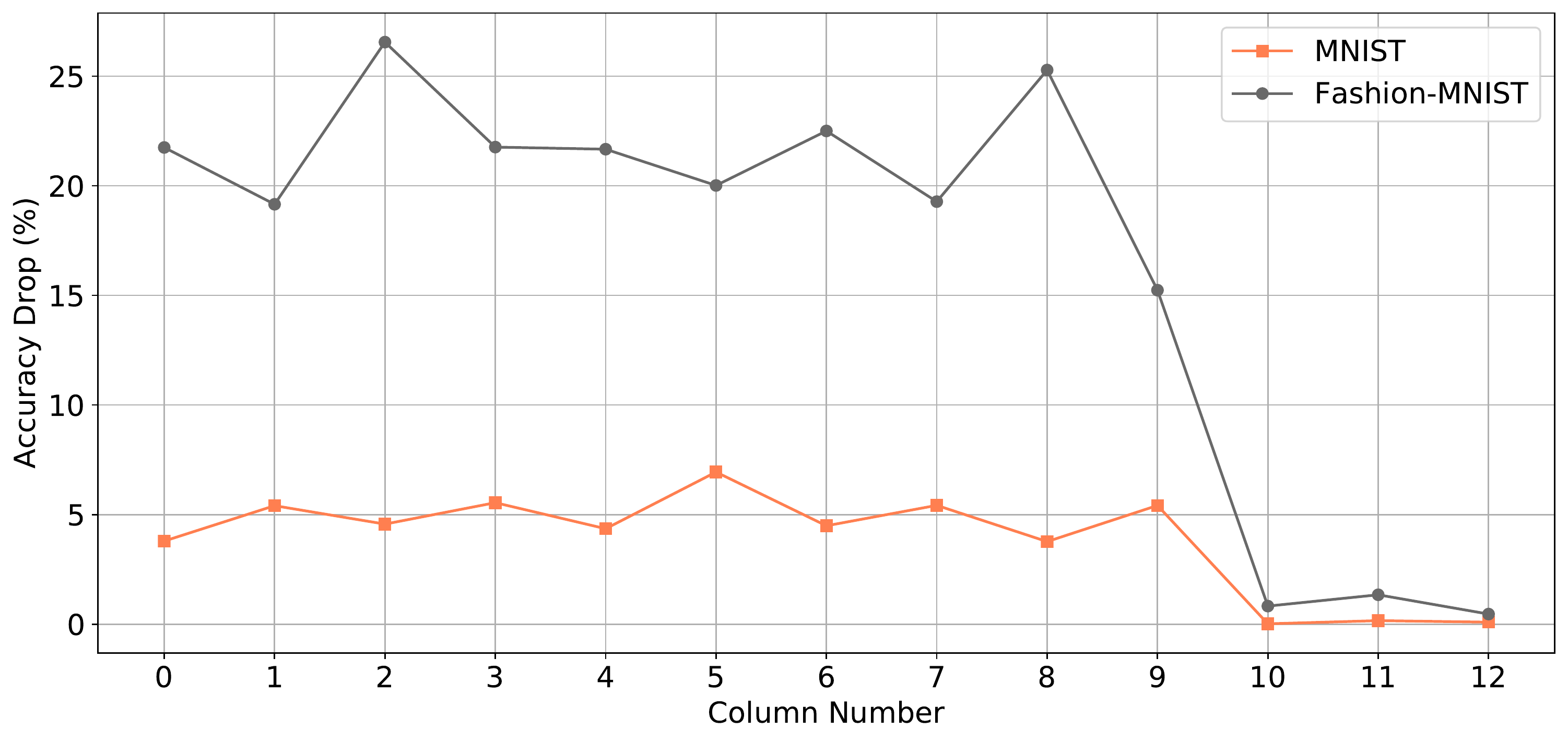}
    \caption{Variation of Accuracy Drop for 20 Faults across Varying Columns of the Weight Matrix.}
    \label{fig:faultLoc}
\end{figure}

Random errors throughout the fault space furnish adequate degradation in classification accuracy. However, the number of random faults required to bring about such degradation is usually quite large. Hence, in this experiment, we focus on estimating the most vulnerable weights in a MLP. \textcolor{black}{The weights corresponding to a particular layer in the MLP are arranged in the form of rows and columns in the 2D DRAM structure, where a particular column of weights corresponds to a specific neuron in the layer \cite{zhang2018analyzing}. Hence, inducing faults along a distinct column is likely to produce a deeper impact on the network performance.}
As described in Section~\ref{subsec:bitpos}, since flipping the sign bit has the most impact on the model performance, we introduce bit-flip errors in the sign bit across only 20 random locations along a particular column of the weight matrix. This erroneous column location is varied, and the corresponding degradation in accuracy, \textcolor{black}{averaged over 10 random runs,} is depicted in Figure~\ref{fig:faultLoc} for both the datasets. We observe that faults in all the columns till column `9' render almost a consistent reduction in accuracy for all the datasets.
However, as the column number exceeds `9', the accuracy drop plummets close to zero, signifying almost negligible impact of the faults on the accuracy of the network. Since the output layer of the network consists of 10 neurons corresponding to 10 distinct classes in all the datasets, the last layer weights are mapped onto the first 10 columns of the matrix. When faults are induced in one among those 10 columns, the computation for that column corresponding to a specific neuron, and thereby to a particular application class is disrupted. With datasets having higher number of classes, such faults can impact differently, affecting the computation in other columns of the systolic array.

Therefore, by injecting bit-wise faults engendering from device-level non-idealities in the primary memory subsystem, we analyzed the reliability of resource-constrained DNN accelerators. We perform an extensive fault characterization of a neural network architecture on multivariate exhaustive datasets. An application-level analysis on the quantized pre-trained inference networks demonstrate degradation of classification accuracy, even at infinitesimal error rates. Hence, it is highly imperative to develop mitigation strategies that protect the most vulnerable network parameters in the memory, in order to improve the performance of the DNN accelerator at sub-optimal DRAM refresh rates.

\section{AI/Deep Learning Accelerator Faults and Performance Impact}
\begin{figure}[h!]
	\centering
	\includegraphics[scale=0.7]{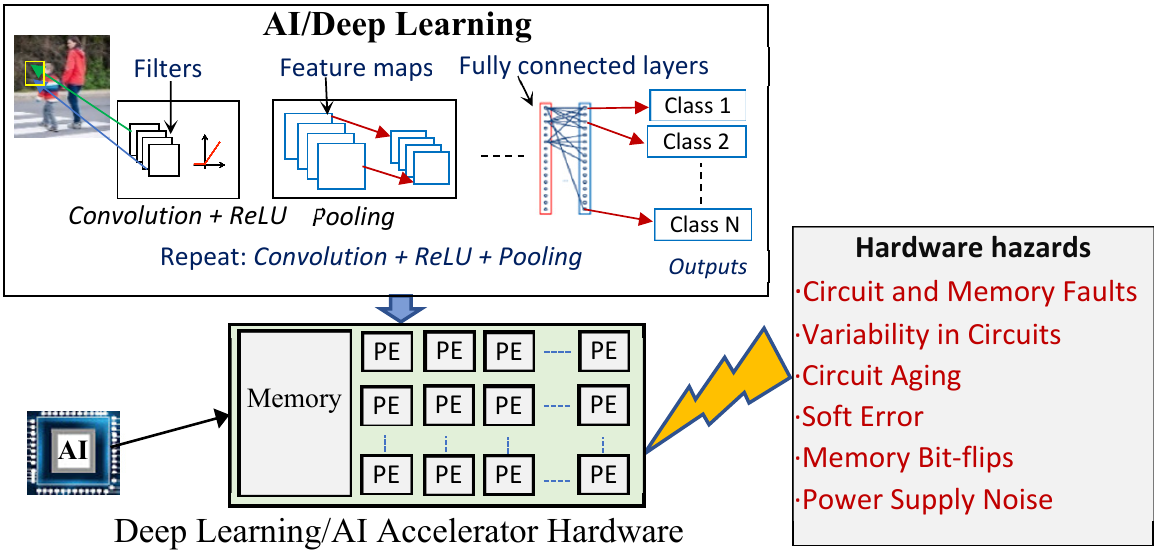}
	\caption {Hardware hazards and faults impacting AI accelerator chips.}
	\label{fig:overview}
\end{figure}

AI and Deep Learning tasks are compute-intensive and require millions of Multiply and Accumulate (MAC) operations for training and inference. In Table \ref{table:mac}, the number of MACs required for classifying a single image from ImageNet benchmark \cite{img} during inference is shown. To increase the execution throughput, the AI accelerators are designed with thousands of densely packed Processing Elements(PEs)/MAC circuits \cite{jouppi2017datacenter}. As a result of this dense integration at advanced technology nodes which are susceptible to defect and yield issues, the AI chips are especially vulnerable to circuit faults (Fig. \ref{fig:overview}).

\begin{table}[]
\centering
\caption {Multiplication and additions in CNNs to classify one image \cite{tcad_yield21}}
\begin{tabular}{|c|c|c|c|c|}
\hline
\begin{tabular}[c]{@{}c@{}}CNN\\ Architecture\end{tabular} & \begin{tabular}[c]{@{}c@{}}Conv2d\\ Layers\end{tabular} & \begin{tabular}[c]{@{}c@{}}Linear\\ Layers\end{tabular} & \begin{tabular}[c]{@{}c@{}}Number of\\ Multiplications\end{tabular} & \begin{tabular}[c]{@{}c@{}}Number of\\ Additions\end{tabular} \\ \hline
LeNet-5                                                    & 3                                                       & 2                                                       & 416,520                                                             & 416,520                                                       \\ \hline
AlexNet                                                    & 5                                                       & 3                                                       & 714,188,480                                                         & 714,188,480                                                   \\ \hline
VGG-16                                                     & 13                                                      & 3                                                       & 15,470,264,320                                                      & 15,470,264,320                                                \\ \hline
ResNet-50                                                  & 53                                                      & 1                                                       & 3,729,522,688                                                       & 1,761,820,672                                                 \\ \hline
\end{tabular}
\label{table:mac}
\end{table}

\subsection{Circuit and Hardware Hazards on Deep Learning/AI Accelerator Fidelity}
The major types of circuit and transistor level hazards that can impact the performance of AI/Deep Learning accelerator are,  (i) process variation induced circuit parameter variations, (ii) runtime power supply voltage noise and droop, (iii) circuit aging with time, and (iv) radiation induced soft errors. For safety-critical applications, chips with permanent stuck-at faults are generally discarded according to defective parts per million/billion (DPPM/DPPB) guidelines of FuSa standards\cite{iso26262}. However, aging induced in-field stuck-at faults can be a concern for automotives.

\subsubsection{Process Variations:} Process  parameter variations are caused by semiconductor manufacturing imperfections, which is a major concern for current advanced technologies (e.g., 10nm and newer), and impact circuit performance at both the memory and MAC units in accelerator hardware. As a result, different samples of the same accelerator chip might exhibit different frequencies (i.e., speed binning \cite{bini}) post-fabrication. For safety-critical AV application, it is important to analyze how process variations can impact the Deep Learning accuracy and fidelity. Unlike other areas, a one-size-fits-all training method may not be suitable for Deep Learning used in safety-critical domains.

\subsubsection{Power Supply Noise:} Voltage noise is caused by simultaneous switching events inside the chip. Since the accelerator chips will perform millions of operations to infer decisions from analysis of the images, the extensive switching inside the MAC and memory units from this can cause voltage noise \cite{PSN}. The corresponding transient voltage droop can cause timing violations or bit-flips inside the accelerator.

\subsubsection{Circuit Aging:} Circuit aging is a critical reliability problem in modern VLSI chips \cite{BIST}. Since aging is use-case dependent and cannot be accurately estimated at time 0, it is a major concern for safety-critical applications. Aging can impact the weight storage SRAM modules by altering their read/write stability with time, and thus cause bit-flip errors.  The Deep Learning accelerator's performance (i.e., operating frequency $F_{MAX}$) might shift with time depending on the amount of usage. 

\subsubsection {Soft Errors:} Although most of the AI accelerators are used at sea-level altitude, it may seem they are immune to soft errors caused by high-energy neutrons from cosmic radiation. However, as shown in detail in \cite{error_prop_ser}, because of weight reuse in Deep Learning, soft-errors can indeed impact the accuracy of accelerators. Hence, it is essential to embed resiliency and robustness against random bit-flips in addition to radiation hardening of the critical MSB bits.

The key modules of accelerators that are susceptible to defects are the weight storage SRAM/Register Files (RF) and the PE/MACs. For SRAM/RF, the faults are generally repaired with ECC and spare cells. The timing-faults in MAC can be solved by appropriate timing guard-band and run-time frequency adjustment. The stuck-at faults in the MAC are permanent and can adversely impact accuracy. In this section accuracy impact of stuck-at faults in MAC are analyzed.

\begin{figure}[h!]
	\centering
	\includegraphics[scale=0.75]{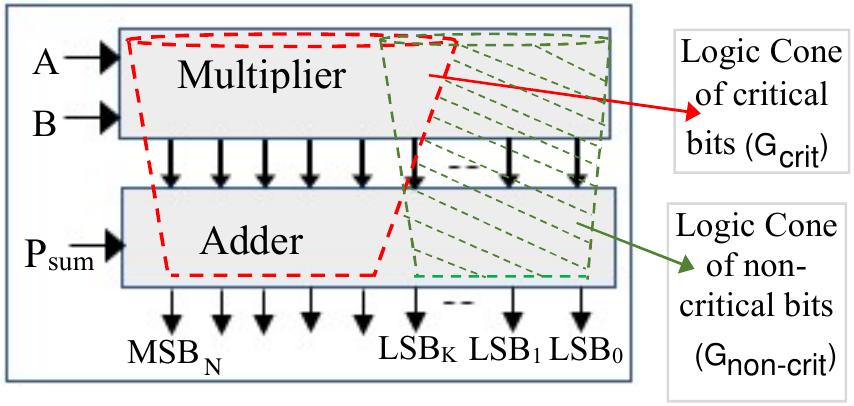}
	\caption {Faults are classified as critical or non-critical based on their location in the logic cone of the bits.}
	\label{fig:logiccone}
\end{figure}

\begin{figure}[h!]
	\centering
	\includegraphics[scale=0.68]{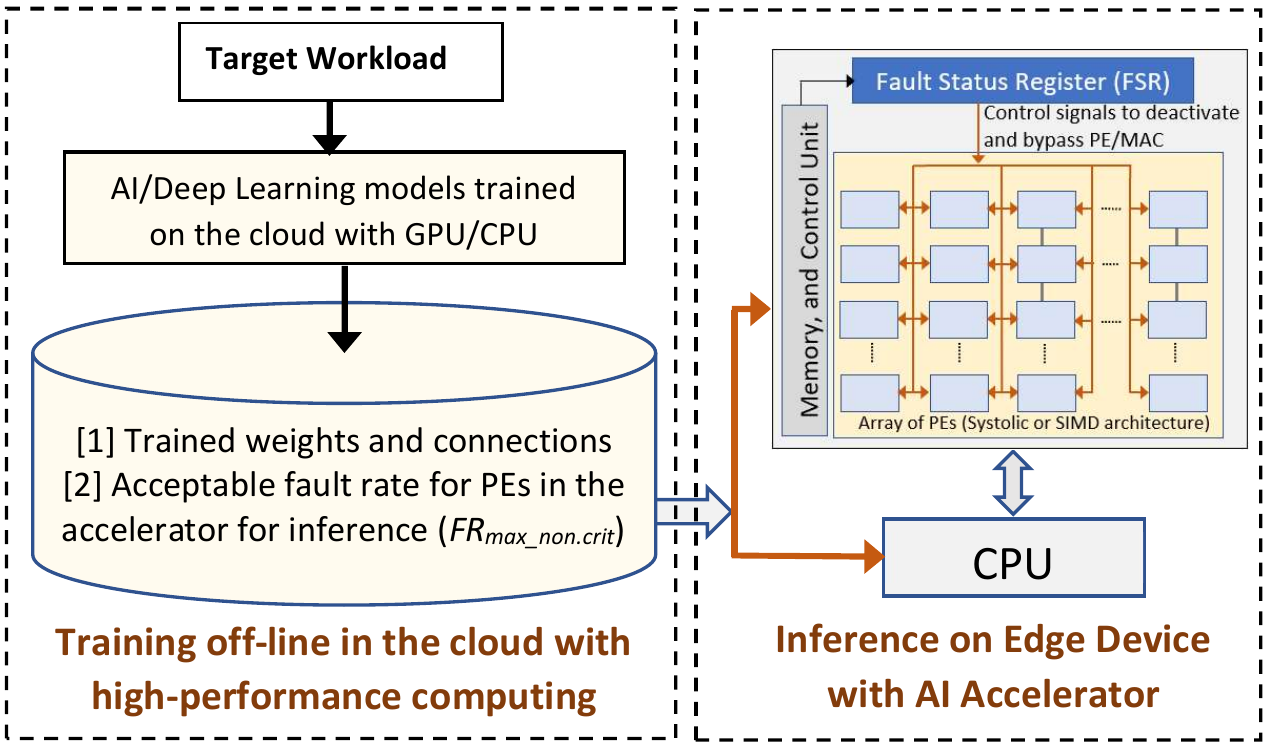}
	\caption {Deactivating some non-critically faulty PEs to keep fault rate within $FR_{max\_non-crit}$ to ensure accuracy does not degrade beyond acceptable limit.}
	\label{fig:faultdeact}
\end{figure}

\begin{figure}[h!]
	\centering
	\includegraphics[scale=1.15]{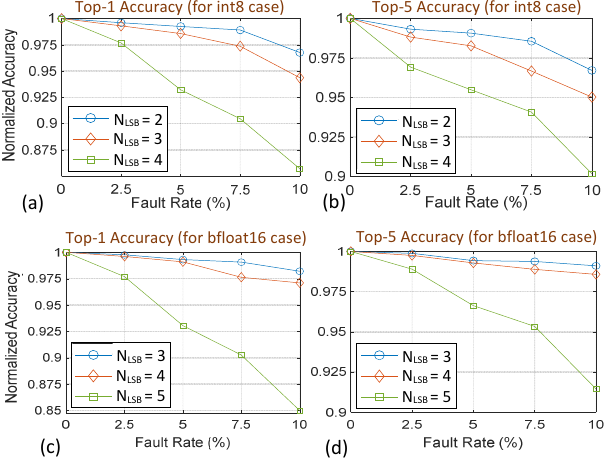}
	\caption {Impact of faults in logic cones of LSB positions  on inference accuracy for AlexNet with ImageNet data set. (a), (b) int8 data format and MAC type.; (c), (d) bfloat16 data format and MAC type.}
	\label{fig:k_vary}
\end{figure}

\begin{figure*}[h]
	\centering
	\includegraphics[width=7.1in,height=2.6in]{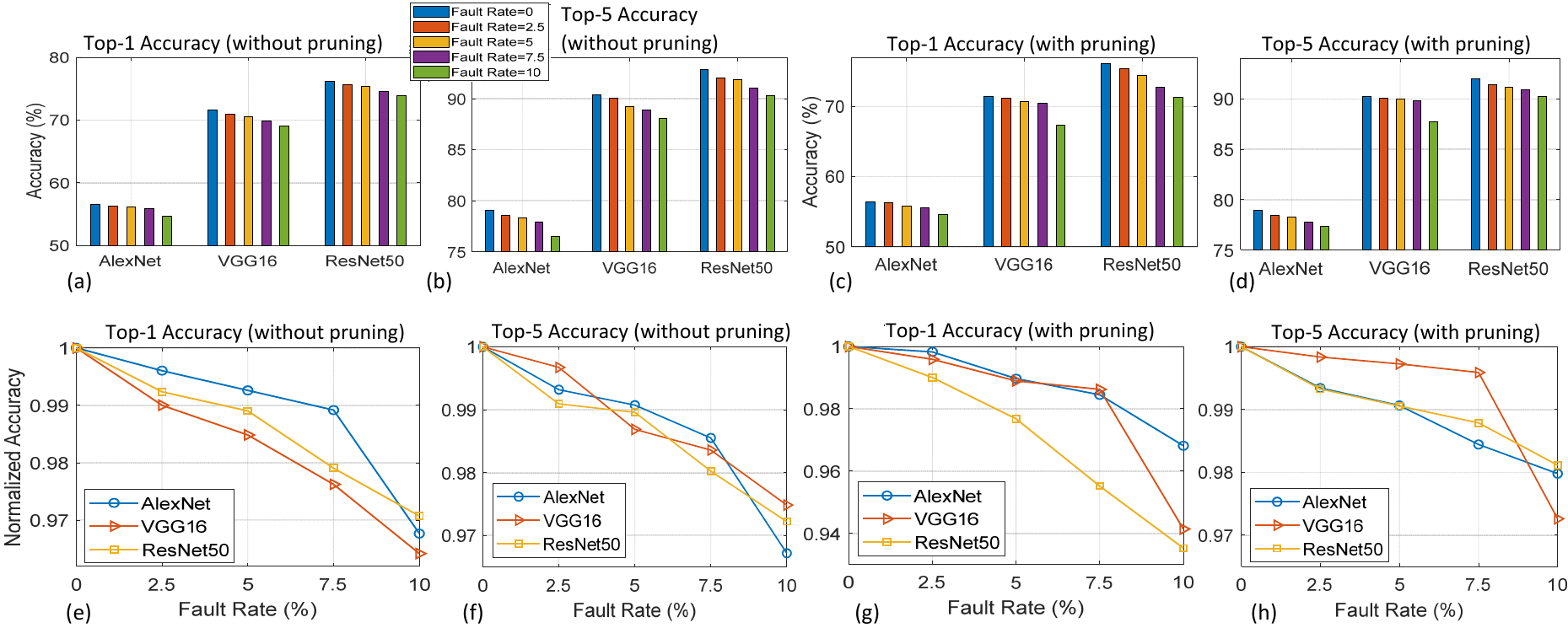}
	\caption {Inference accuracy changes with fault (non-critical) rates for CNNs running imagenet dataset. (a), (b) Top-1 and Top-5 accuracy changes ;  (c), (d) Top-1 and Top-5 accuracy changes with model pruning; (e) to (h) normalized accuracy changes in Top-1 and Top-5 (with and without pruning) for fault rates.}
	\label{fig:cnn}
\end{figure*}

\subsection {Fault Location and Accuracy Impact:}
Systematic defects and yield losses are caused by layout-sensitive lithographic hotspots and other process imperfections, variations, and are generally independent of the layout area. On the other hand, random defect generated yield losses are caused by defect particles and are dependent on the standard-cell or the layout area as well as the defect particle size \cite{yield}. These defects (e.g., short/open defect, poor contact/via, etc.) and corresponding circuit faults can occur at different sites inside the MAC circuit block. The precision loss - due to the presence of circuit faults - at the output of multiply and accumulate operation will depend on the location of the fault inside the MAC circuit and the logic cone impacted by the fault as shown in Fig. \ref{fig:logiccone}. For example, if a multiplier circuit that performs the multiplication of two 8-bit numbers, has faults impacting up to $K$ LSB bits, then it will sustain a worst-case error of $\pm\sum_{i=0}^{K+1}2^i$ (the last $2^{K+1}$ term comes from the worst-case carry-in path of partial product addition, as explained in the next subsection). As $K$ increases, the faults impact the more significant digits causing the worst-case error of the multiplier to increase. Errors will also occur in the adder circuit of the MAC if it is corrupted by faults. Since the multiplier is the more area-dominant block in a MAC, it will be more prone to faults and computation errors. A similar analysis can be also done for bfloat16 \cite{bfloat} MAC to identify faults in LSB positions (i.e., logic cone of corresponding bits) and their impact on accuracy. If any PE has a fault in the critical logic cone then it needs to be deactivated to prevent accuracy loss because critical faults can cause large magnitude error in MAC output. For faults in the non-critical logic cone, up to a certain fault rate (i.e., $FR_{max\_non-crit}$) may be acceptable.

The IDs of faulty PEs and their fault type (e.g., critical or non-critical) can be identified with ATPG test patterns and recorded in a Fault Status Register (FSR) \cite{tcad_yield21}. During inference, the mobile/edge accelerator's FSR and control unit reads the $FR_{max\_non-crit}$ and if it is lower than the current fault rate $FR_{non-crit}$, then the control unit sends deactivation signals to disable some of the PEs. The deactivation protocol and the complete map of faulty PE locations are programmed in firmware/software by the manufacturer. With the user-given input of acceptable fault rate and the stored PE fault map, the protocol will automatically disable a few faulty PEs  to ensure that the overall faulty PE rate of the accelerator does not exceed a threshold (i.e., $FR_{max\_non-crit}$). When deactivating some faulty PEs, the firmware/software will ensure that the remaining faulty PEs are not clustered. As shown in Fig. \ref{fig:faultdeact}, control signals are transmitted from the FSR to all the PEs to selectively disable the faulty PEs when needed. By adopting this scheme, the manufacturer can avoid discarding the full accelerator chip/die only because of the presence of few PEs with faulty MACs, and thereby increase yield. The overhead in this yield loss reduction are the extra on-chip register (FSR) to store the IDs and the control signal routes to disable faulty PEs \cite{tcad_yield21}.

\begin{figure}[h!]
	\centering
	\includegraphics[scale=0.9]{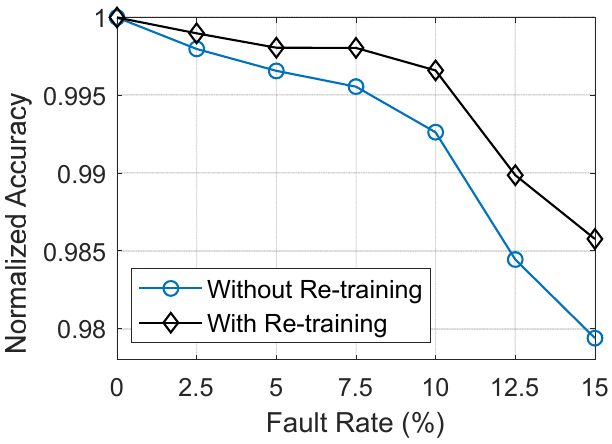}
	\caption {Improvement in accuracy with fault-aware training on LeNet-5.}
	\label{fig:retr}
\end{figure}

\subsection{Accelerator Architecture and Faulty PE Deactivation}
AI accelerators can be designed either in SIMD or Systolic architectures. In SIMD architecture the loosely coupled independent Processing Elements (PEs) are connected with NoC or mesh and can be individually switched off and bypassed with wires \cite{chen2016eyeriss,jouppi2017datacenter,tcad_yield21}. The tightly-coupled 2D systolic-arrays in TPU introduce challenges in deactivating and bypassing individual PEs. A common technique is to use spare PE blocks to substitute for faulty PEs. However, this approach requires complex wiring between spare and faulty PEs \cite{FT_syst}. An innovative software-level technique to deactivate and bypass faulty PEs in systolic array without any hardware-level modifications was proposed in \cite{tcad_yield21}.

\subsection{Results and Analysis}
To identify the impact of the number of LSBs having faults in their logic cones, the number of LSBs were varied from 2 to 4 for int8 and 3 to 5 for bfloat16 \cite{bfloat} data formats for the CNN model AlexNet with the ImageNet test set. These experiments were done with PyTorch. The results are shown in Fig. \ref{fig:k_vary}, where X-axis is fault-rate in non-critical logic cones (the LSBs). From this analysis, we conservatively selected 2 LSBs for int8 and 4 mantissa LSBs for the bfloat16 hardware models as non-critical (i.e., circuit faults occurring in their logic cones are non-critical), and rest of all the bits are considered critical. Any PE with one or more critical faults are considered defective and must be deactivated to ensure fidelity in accuracy.  In the experiments the accelerator hardware comprised of 128 by 128 array of PE/MACs. The faulty PEs are modeled as uniformly distributed across the accelerator columns with fault probability of $FR\%$, the fault rate.  This implies that for $FR\%$ fault rate, each column of the accelerator has $0.01*FR*N_{Row}$ faulty PEs randomly distributed across that column.

The accuracy changes - in the standard Top-1 and Top-5 format - with MAC faults are shown in Fig. \ref{fig:cnn} for 50,000 test images from ImageNet \cite{img}. We observe from Fig. \ref{fig:cnn} that up to a certain fault (non-critical)  rate might be acceptable depending on the desired accuracy of the AI/Deep Learning task.

\subsubsection{Fault-aware training to enhance Robustness}
Using a fault-aware training flow some of the accuracy loss due to faults in MAC units can be recovered by incorporating the fault effects in the backpropagation-based weight update segment and allowing the DNN to adapt accordingly \cite{tcad_yield21}. To experimentally demonstrate this technique, we used the LeNet-5 CNN architecture. Results from this fault-aware training are shown in Fig. \ref{fig:retr}. For 7.5\% fault rate (non-critical faults) the normalized accuracy loss improved from 0.5\% to 0.22\% due to fault-aware training.

\section{Dependability Issues in Neuromorphic Computing}
\begin{figure}[h!]
	\centering
	\centerline{\includegraphics[width=0.79\columnwidth]{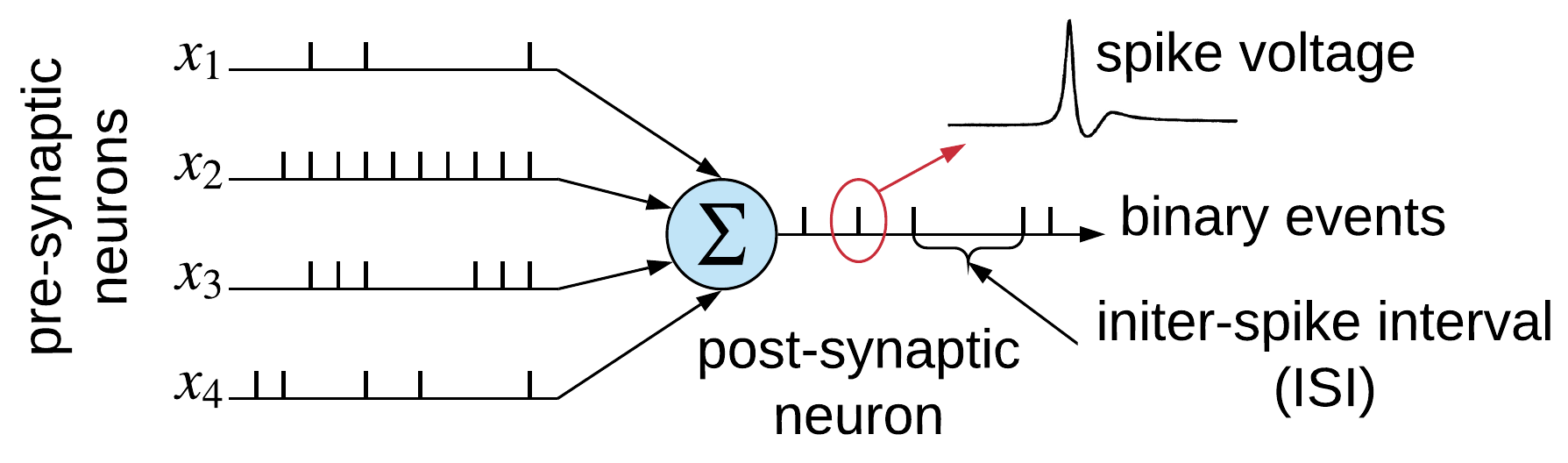}}
	\caption{Integration of spike trains at the post-synaptic neuron from four pre-synaptic neurons in a Spiking Neural Network (SNN). Each spike is a voltage waveform of small time duration to the order of ms.}
	\label{fig:snn}
\end{figure}

Neuromorphic Computing is a term coined by Carver Mead in the late 1980s describing Very Large-Scale Integration (VLSI) systems, which mimic the neuro-biological architecture of the central nervous system~\cite{mead1990neuromorphic}.
Neuromorphic systems are energy efficient in executing Spiking Neural Networks (SNNs), which are considered as the third generation of neural networks. SNNs use spike-based computations and bio-inspired learning algorithms in solving pattern recognition problems~\cite{maass1997networks}. In an SNN, pre-synaptic neurons communicate information encoded in spike trains to post-synaptic neurons, via synapses. Performance, e.g., accuracy of an SNN model can be assessed in terms of the inter-spike interval (ISI), which is defined as inverse of the mean firing rate of neurons. This is illustrated in Figure~\ref{fig:snn}.

Recently, neuromorphic platforms such as DYNAPs~\cite{dynapse}, TrueNorth~\cite{truenorth}, and Loihi~\cite{loihi} are introduced to the systems community. These systems are designed as tile-based architectures with a shared interconnect for communication~\cite{balaji2019exploration,balaji2019design,catthoor2018very}. 
A tile may consist of a crossbar for mapping neurons and synapses of an application. 
Recently, Non-Volatile Memory (NVM) 
technologies 
such as Phase-Change Memory (PCM) and Oxide-based Resistive RAM (OxRRAM) 
are used to implement 
synaptic storage in each crossbar~\cite{Mallik2017,Burr2017}.\footnote{Beside neuromorphic computing, some of these NVM technologies are also used to implement main memory in conventional computers to improve performance and energy efficiency~\cite{palp,mneme,datacon,hebe,song2020design}.} 
NVMs bring certain advantages such as high integration density, multi-bit synapses, CMOS compatibility, and above all, non-volatility, which can further lower the energy consumption of neuromorphic hardware. However, NVMs also introduce reliability issues such as endurance, aging, and read disturbances (see Table~\ref{tab:reliability_summary} for a summary of these issues). In this paper, we will discuss some of these issues and our approach to mitigate them.

\begin{table}[h]
\renewcommand{\arraystretch}{1.2}
\setlength{\tabcolsep}{2pt}
\caption{Reliability issues in NVMs.}
\vspace{-5pt}
\label{tab:reliability_summary}
\centering
{\fontsize{10}{10}\selectfont
\begin{tabular}{|l|c|}
\hline
\textbf{Reliability Issues} & \textbf{NVMs}\\
\hline
High-voltage related circuit aging & PCM, Flash\\
High-current related circuit aging & OxRAM, STT-MRAM\\
Read disturbance & All\\
Limited endurance & All\\
\hline
\end{tabular}}
\end{table}

\subsection{Introduction to Non-Volatile Memory}
Emerging NVM technologies such as phase-change memory (PCM), oxide-based memory (OxRAM), spin-based magnetic memory (STT-MRAM), and Flash have recently been used to implement synaptic storage in neuromorphic hardware. NVMs are non-volatile, have high CMOS compatibility, and can achieve high integration density. Each NVM device can implement both a single-bit and multi-bit synapse. 
Because of these properties, an NVM-based neuromorphic hardware typically consumes energy that is in the order of magnitudes lower than using SRAMs~\cite{Mallik2017}.

Without loss of generality, we discuss the dependability issues for PCM-based neuromorphic computing. PCM is one of the mature NVM technologies and is successfully demonstrated in recent neuromorphic prototypes~\cite{Burr2017}.

\begin{figure}[h!]
	\centering
	\vspace{-10pt}
	\centerline{\includegraphics[width=0.99\columnwidth]{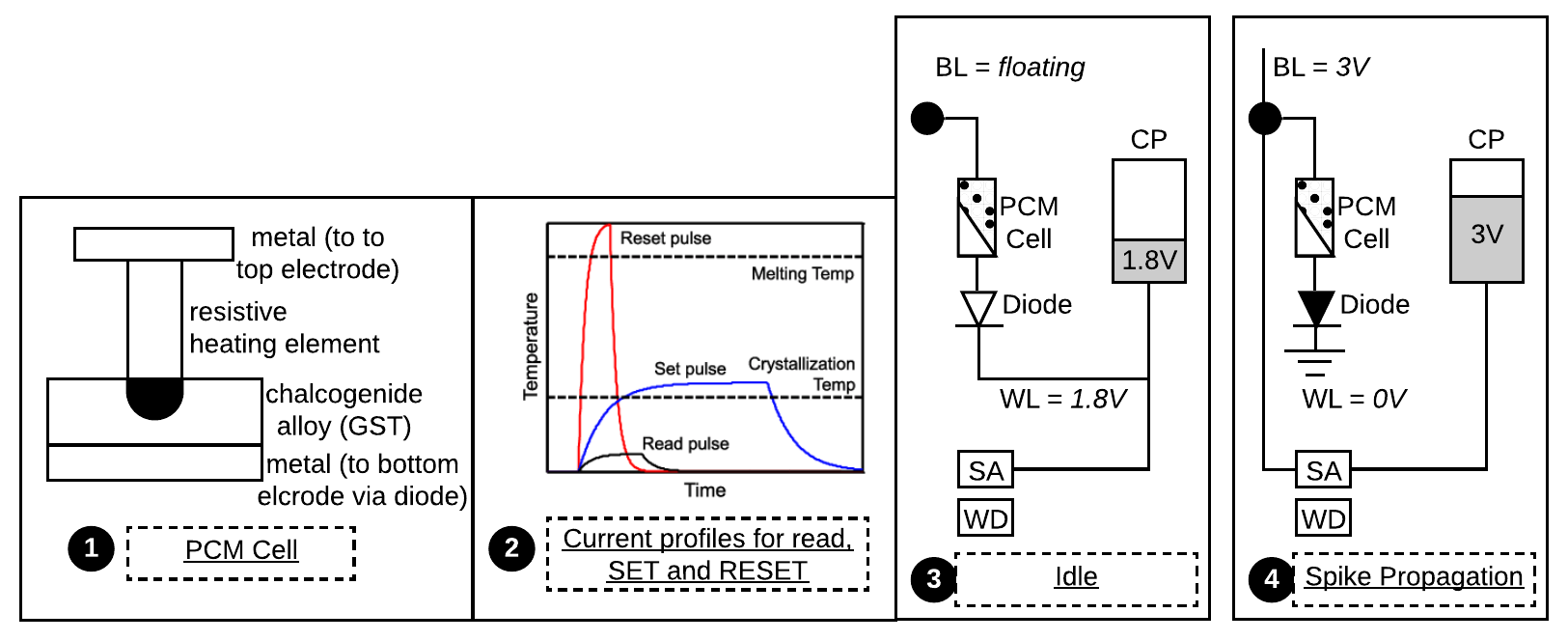}}
	\vspace{-5pt}
	\caption{Operation of PCM in neuromorphic computing.}
	\vspace{-10pt}
	\label{fig:pcm}
\end{figure}

Figure~\ref{fig:pcm}~\ding{182} illustrates how a chalcogenide semiconductor alloy is used to build a PCM cell.
The amorphous phase (RESET) in this alloy has higher resistance than the crystalline phase (SET).
Ge${}_2$Sb${}_2$Te${}_5$ (GST) is the most commonly used alloy for PCM.
To compute \ineq{(x_i\cdot w_i)}, a current is injected into the resistor-chalcogenide junction via the heater element. The current is controlled to ensure that the phase of the PCM cell is not disturbed. This is the fundamental operation of forward propagation of neuron excitation during inference. For online learning (e.g., using STDP), the injected current induces (Joule) heating in the chalcogenide alloy, changing its conductivity, thereby achieving synaptic weight updates. 
Figure~\ref{fig:pcm}~\ding{183} illustrates the current profiles necessary for inference (using the read pulse) and online learning (using SET and RESET pulses) in PCM. 
These current profiles are generated using an on-chip charge pump (CP).
Figure~\ref{fig:pcm}~\ding{184} illustrates the PCM cell's operation when idle, i.e., when a neuron is not activated. We illustrate a 1D-1R structure, where a single PCM cell is connected to a row and column using a diode as an access device. Diode-based PCM cells allow very high integration density in scaled technology nodes compared to transistor-based PCM. The CP is operated at 1.8V to maintain the required biasing. Finally, Figure~\ref{fig:pcm}~\ding{185} illustrates the PCM operation during inference. The CP is operated at 3V to generate the read current profile of Figure~\ref{fig:pcm} \ding{183} using the sense amplifier (SA). The write driver (WD) is used for generating the currents for online learning.

These high-voltage operations of the charge pump (and the peripheral circuit of a crossbar) accelerate circuit aging, lowering the dependability of neuromorphic computing.

Apart from diode, transistors are also used as access devices. When using transistor-based PCM cells, the CP is operated at lower voltages: 1.2V during idle and 1.8V during spike propagation. Though aging issues are less severe in such designs, they are still a dependability concern for neuromorphic computing.

Peripheral aging are not the only dependability issues in neuromorphic hardware. 
Unfortunately, NVMs have limited endurance, ranging from
\ineq{10^5} (for Flash) to \ineq{10^{10}} (for OxRRAM), with PCM somewhere in between
(\ineq{\approx 10^{7}}). 

In the following we describe our recent efforts in mitigating aging and endurance in neuromorphic computing.

\subsection{Mitigating Peripheral Aging in Neuromorphic Hardware}
In our prior work~\cite{frameworkCAL,reneu,NeuromorphicLR}, we have analyzed different aging issues in the peripheral circuit of a neuromorphic hardware. We briefly elaborate on these issues.

We consider the CMOS aging due to 
Time-Dependent Dielectric Breakdown (TDDB), Bias Temperature Instability (BTI), and Hot-Carrier Injection (HCI) failure mechanisms. These are the dominant ones in scaled technology nodes (45nm and below). In older nodes, Electromigration (EM) also plays a key role~\cite{das2015reliability,das2013aging,das2014combined,das2013reliability,das2018reliable,das2012fault,das2016slowdown,das2013energy,das2015workload,das2014reinforcement,das2013improving,santos2014criticality,das2014communication,das2014energy,das2012energy,das2014temperature,das2016adaptive,bolchini2013run}.

CMOS aging is accelerated when the device is \emph{stressed}, i.e., exposed to high overdrive voltages\footnote{Overdrive voltage is defined as the voltage between transistor gate and source ($V_{GS}$) in excess of the threshold voltage ($V_\text{th}$), where $V_\text{th}$ is the minimum voltage required between gate and source to turn the transistor on.}. With this understanding, 
we provide a brief background of these failure mechanisms. 
\begin{itemize}
    \item \textit{TDDB:} This is a failure mechanism in a CMOS device, when the gate oxide breaks down as a result of long-time application of relatively low electric field (as opposed to immediate breakdown, which is caused by strong electric field)~\cite{roussel2018new}. The lifetime of a CMOS device is measured in terms of its \textit{mean time to failure} (\textbf{MTTF}) as
\vspace{-6pt}
\begin{equation}
    \label{eq:MTTF_TDDB}
    \vspace{-6pt}
    \footnotesize \text{MTTF}_\text{TDDB} = A.e^{-\gamma\sqrt{V}},
\end{equation}
where \ineq{A} and \ineq{\gamma} are material-related constants, and \ineq{V} is the overdrive gate voltage of the CMOS device. 
\item \emph{BTI:} This is a failure mechanism in a CMOS device in which positive charges are trapped at the oxide-semiconductor boundary underneath the gate~\cite{gao2017nbti}.
BTI manifests as 1) decrease in drain current and transconductance, and 2) increase in off current and threshold voltage. 
The BTI lifetime of a CMOS device is
\vspace{-6pt}
\begin{equation}
    \label{eq:MTTF_NBTI}
    \vspace{-6pt}
    \footnotesize \text{MTTF}_\text{BTI} = \frac{A}{V^\gamma}e^{\frac{E_a}{KT}},
\end{equation}
where \ineq{A} and \ineq{\gamma} are material-related constants, \ineq{E_a} is the activation energy, \ineq{K} is the Boltzmann constant, \ineq{T} is the temperature, and \ineq{V} is the overdrive gate voltage.
\item \emph{HCI:} This is a failure mechanism in a CMOS device, when a carrier (electron or hole) gains sufficient kinetic energy to overcome the potential barrier of the conducting channel and gets trapped in the gate dielectric, permanently changing the CMOS's switching properties~\cite{wan2019hci}.
\end{itemize}

Unlike the TDDB and BTI failure mechanisms, for which silicon-characterized reliability models are available from foundries, characterized models for HCI failure mechanism are still in development for scaled nodes.

Current methods for qualifying reliability are overly conservative, since they estimate circuit aging considering worst-case operating conditions and unnecessarily constrain performance. 
Recent system-level works on mapping SNN-based applications to neuromorphic hardware, such as~\cite{pycarl,dfsynthesizer,balaji2020ESL,spinemap,balajiISVLSI,das2018dataflow,rtmJSPS,twisha_energy,balaji2020compiling,psopart},
target performance improvement only. They do not consider reliability issues of neuromorphic computing.

To address these limitations, we have designed RENEU~\cite{reneu}, a reliability-oriented approach to map machine learning applications to neuromorphic hardware, with the aim of improving system-wide reliability, without compromising key performance metrics such as execution time of these applications on the hardware. 
Fundamental to RENEU is a novel formulation of the aging of CMOS-based circuits in a neuromorphic hardware considering different failure mechanisms. Using this formulation, RENEU develops a system-wide reliability model which can be used inside a design-space exploration framework involving the mapping of neurons and synapses to the hardware. To this end, RENEU uses an instance of Binary Particle Swarm Optimization (PSO)~\cite{khanesar2007novel} to generate mappings that are Pareto-optimal in terms of performance and reliability.
We evaluate RENEU using SNN-based streaming and non-streaming applications~\cite{HeartEstmNN,das2018heartbeat,HeartClassJolpe,moyer2021motif,moyer2020machine}. We evaluate these applications on a state-of-the-art neuromorphic  hardware with PCM synapses. 

\begin{figure}[h!]
	\centering
	\vspace{-5pt}
	\centerline{\includegraphics[width=0.99\columnwidth]{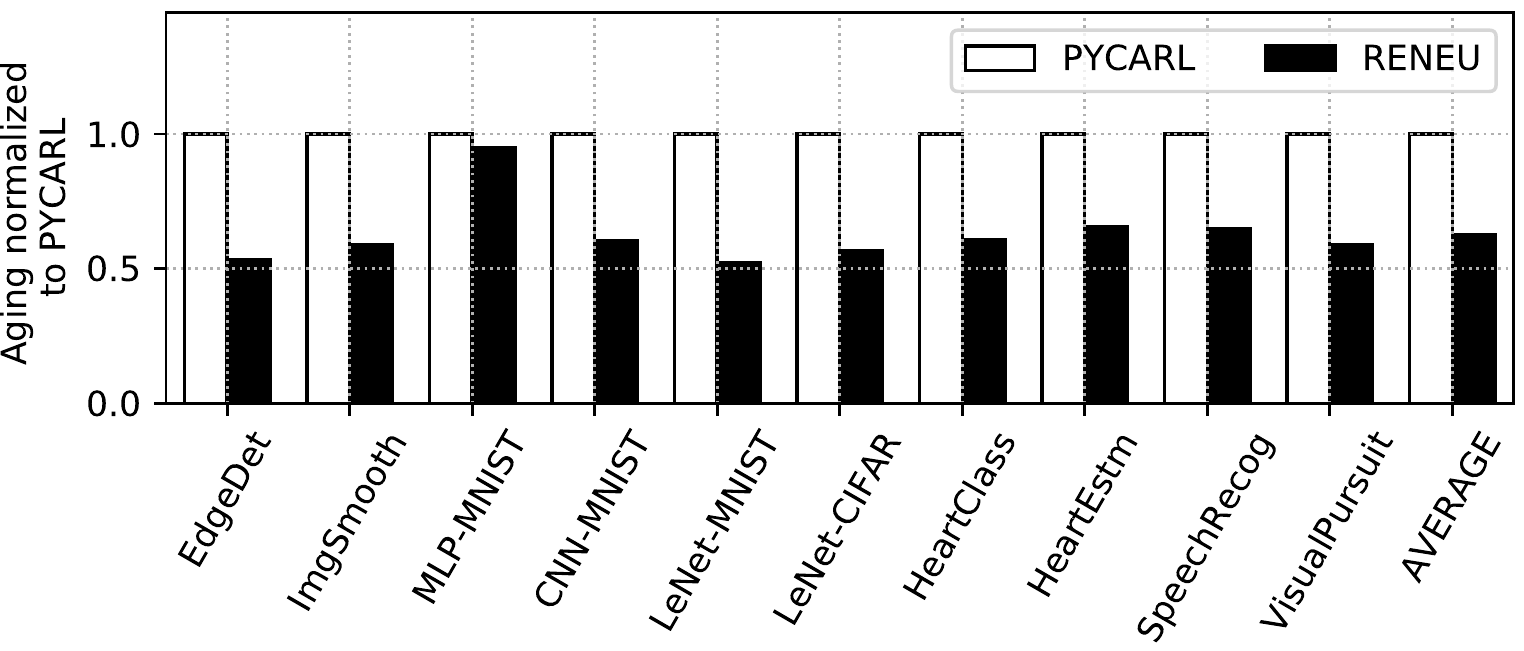}}
	\vspace{-5pt}
	\caption{Circuit aging in the neuromorphic hardware normalized to PyCARL (lower is better)~\cite{reneu}.}
	\vspace{-10pt}
	\label{fig:aging}
\end{figure}

Figure~\ref{fig:aging} reports the circuit aging caused by RENEU normalized to PyCARL, a state-of-the-art SNN mapping approach. We plot results for each of our machine learning applications. 
We observe that the aging of RENEU is lower than PyCARL by an average of 38\%. This improvement is because RENEU formulates the detailed circuit aging of a neuromorphic hardware and allocates the neurons and synapses of a machine learning application to minimize it. 

\subsection{Mitigating Thermal and Endurance Issues in Neuromorphic Hardware}
In our prior work~\cite{twisha_endurance,twisha_thermal,twisha_tpds}, we have formulated the thermal issues in neuromorphic hardware and established its impact on endurance. 

\begin{figure}[h!]
	\centering
	\vspace{-10pt}
	\centerline{\includegraphics[width=0.69\columnwidth]{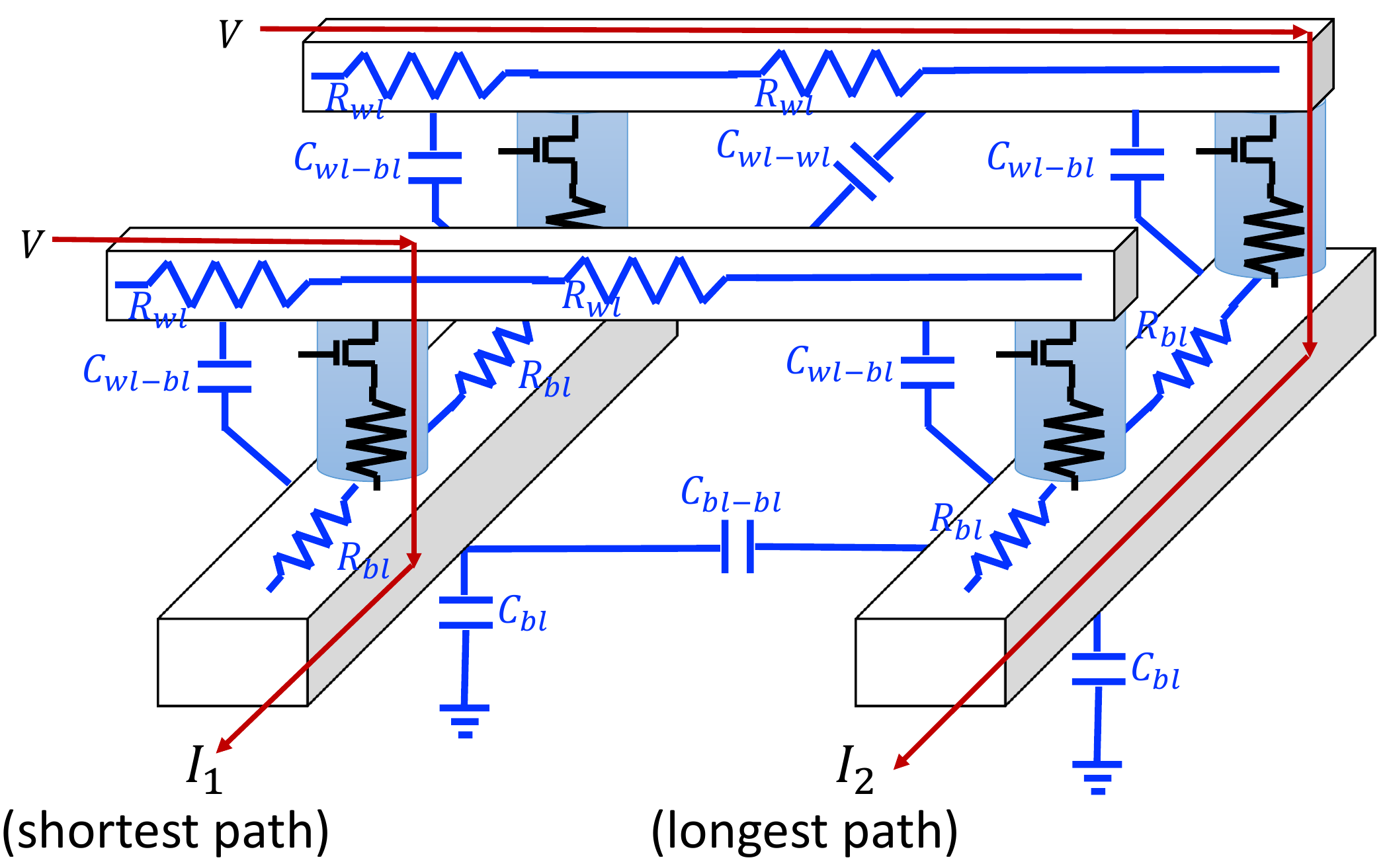}}
	\vspace{-5pt}
	\caption{Bitline and wordline parasitics in a PCM crossbar.}
	\vspace{-10pt}
	\label{fig:parasitics}
\end{figure}

We analyze the internal architecture of a PCM crossbar (see Fig.~\ref{fig:parasitics}) and observe that parasitic components on horizontal and vertical wires of a crossbar are a major source of parasitic voltage drops in the crossbar.
Using detailed circuit simulations at different process (P), voltage (V), and temperature (T) corners, we show that these voltage drops 
create
current variations in the crossbar.
For the same spike voltage, 
current on the shortest path is significantly higher than the current on the longest path in the crossbar, where the length of a current path is measured in terms of its number of parasitic components.
These current variations create
asymmetry in the self-heating temperature of PCM cells during their weight updates, e.g., during model training and continuous online learning~\cite{chen2018lifelong}, which directly influences their endurance.

\begin{figure}[h!]%
    \centering
    \vspace{-10pt}
    \centerline{\includegraphics[width=0.99\columnwidth]{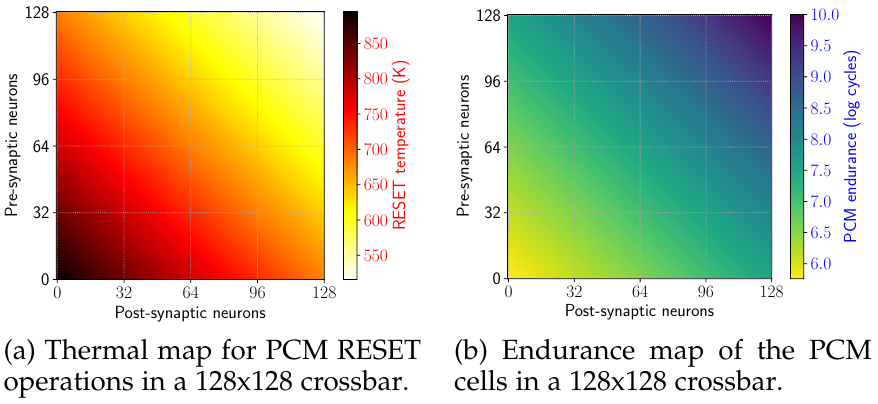}}
	\vspace{-5pt}
    \caption{Temperature and endurance map of a 128x128 PCM crossbar at \ineq{65}nm process node with \ineq{T_{amb} = 298K}.}%
    \label{fig:temperature_endurance_map}%
    \vspace{-10pt}
\end{figure}

Figure~\ref{fig:temperature_endurance_map} plots the temperature and endurance maps of a 128x128 crossbar at {65}nm process node with \ineq{T_{amb} = 298K}. The PCM cells at the bottom-left corner have higher self-heating temperature than at the top-right corner. This asymmetry in the self-heating temperature creates a wide distribution of endurance, ranging from \ineq{10^6} cycles for PCM cells at the bottom-left corner to \ineq{10^{10}} cycles at the top-right corner. These endurance values are consistent with the values reported for recent PCM chips from IBM~\cite{burr2016recent}.

Using such technology modeling, we propose eSpine, a novel technique to improve endurance-related lifetime of PCM by incorporating the endurance variation within each crossbar in mapping machine learning workloads, ensuring that synapses with higher activation are always implemented on PCM cells with higher endurance, and vice versa. eSpine works in two steps. First, it uses the Kernighan-Lin Graph Partitioning algorithm to partition a workload into clusters of neurons and synapses, where each cluster can fit in a crossbar. Second, it uses an instance of PSO to map clusters to tiles, where the placement of synapses of a cluster to the PCM cells of a crossbar is performed 
by analyzing their activation within a workload.
We evaluate eSpine for a state-of-the-art neuromorphic hardware model with PCM synapses. Using 10 SNN workloads, 
we
demonstrate a significant improvement in the effective lifetime.

\section{Conclusion} \label{sec:conclusion}

To conclude, as process technology continues to scale aggressively, reliability issues in neuromorphic hardware are becoming a primary concern for system developer. First, we analyzed the reliability of DNN accelerators in the presence of DRAM faults. By injecting bit-wise faults engendering from device-level non-idealities in the memory subsystem of the accelerator, we perform an extensive fault characterization of multiple DNN architectures on multivariate exhaustive datasets. An application-level analysis on the quantized pre-trained inference networks demonstrate degradation of classification accuracy, even at infinitesimal error rates. Next, we analyzed the MAC circuits, as they occupy a significant portion of the DNN hardware. Circuit faults in MSB logic cones of the MAC can adversely impact accuracy. For MACs with faults in LSB logic cones, the fraction of such faulty MACs must be within an application-dependent range to ensure accuracy degradation does not exceed acceptable limits.  System-level approaches, such as the ones we highlighted in this paper, mitigate these reliability issues via intelligent neuron and synapses placement on the hardware. These approaches, however, can be further improved by incorporating technology perspective within a holistic design-technology co-optimization framework. 

\balance 
\bibliographystyle{IEEEtran}
\bibliography{commands,references,disco}

\end{document}